\documentclass[
reprint, 
superscriptaddress,
bibnotes,
amsmath,amssymb,
aps,
prl,
floatfix
]{revtex4-2}

\usepackage{dcolumn}
\usepackage{bm}

\usepackage[dvipdfmx]{graphicx}
\usepackage{color}
\usepackage{footnote}
\usepackage{lineno}
\usepackage{amsmath}
\usepackage{comment}
\usepackage{siunitx}
\DeclareSIUnit[number-unit-product = ]\percent{\%} 
\usepackage{mathtools}
\usepackage[caption=false]{subfig} 

\newcommand{\taun}{$\tau_{\mathrm{n}}$}
\newcommand{\HeT}{$^3$He}
\newcommand{\Hecapture}{$^3$He(n,p)$^3$H}
\newcommand{\Ncapture}{$^{14}$N(n,p)$^{14}$C}
\newcommand{\Ocapture}{$^{17}$O(n,$\alpha$)$^{14}$C}
\newcommand{\Ccapture}{$^{12}$C(n,$\gamma$)$^{13}$C}

\newcommand{\LiF}{$^6$LiF tile}

\begin{document}
\date{\today}
\title{Improved measurements of neutron lifetime with cold neutron beam at J-PARC}

\author{Y.~Fuwa}
\affiliation{Japan Atomic Energy Agency, Tokai 319-1195, Japan}
\affiliation{J-PARC Center, Tokai 319-1195, Japan}

\author{T.~Hasegawa}
\affiliation{Department of Physics, Graduate School of Science, Nagoya University, Nagoya 464-8602, Japan}

\author{K.~Hirota}
\affiliation{High Energy Accelerator Research Organization, Tsukuba 305-0801, Japan}

\author{T.~Hoshino}
\affiliation{Department of Physics, Graduate School of Science, Kyushu University, Fukuoka 819-0395, Japan}

\author{R.~Hosokawa}
\affiliation{Faculty of Science and Engineering, Iwate University, Morioka  020-8551, Japan}

\author{G.~Ichikawa}
\affiliation{High Energy Accelerator Research Organization, Tsukuba 305-0801, Japan}
\affiliation{J-PARC Center, Tokai 319-1195, Japan}

\author{S.~Ieki}
\affiliation{Department of Physics, Graduate School of Science, The University of Tokyo, Tokyo 113-0033, Japan}

\author{T.~Ino}
\affiliation{High Energy Accelerator Research Organization, Tsukuba 305-0801, Japan}
\affiliation{J-PARC Center, Tokai 319-1195, Japan}

\author{Y.~Iwashita}
\affiliation{Institute for Integrated Radiation and Nuclear Science, Kyoto University, Osaka 590-0494, Japan}

\author{M.~Kitaguchi}
\affiliation{Department of Physics, Graduate School of Science, Nagoya University, Nagoya 464-8602, Japan}
\affiliation{Kobayashi-Maskawa Institute, Nagoya University, Nagoya 464-8602, Japan}

\author{R.~Kitahara}
\affiliation{Department of Physics, Kyoto University, Kyoto, 606-8502, Japan}
\affiliation{Technology Research Center, Sumitomo Heavy Industries, Ltd., Yokosuka,
237-8555, Japan}

\author{S.~Makise}
\affiliation{Department of Physics, Graduate School of Science, Kyushu University, Fukuoka 819-0395, Japan}

\author{K.~Mishima}
\email{Corresponding author: kmishima@kmi.nagoya-u.ac.jp}
\affiliation{Kobayashi-Maskawa Institute, Nagoya University, Nagoya 464-8602, Japan}
\affiliation{High Energy Accelerator Research Organization, Tsukuba 305-0801, Japan}

\author{T.~Mogi}
\affiliation{Department of Physics, Graduate School of Science, The University of Tokyo, Tokyo 113-0033, Japan}

\author{N.~Nagakura}
\affiliation{Department of Physics, Graduate School of Science, The University of Tokyo, Tokyo 113-0033, Japan}

\author{H.~Oide}
\affiliation{High Energy Accelerator Research Organization, Tsukuba 305-0801, Japan}

\author{H.~Okabe}
\affiliation{Department of Physics, Graduate School of Science, Nagoya University, Nagoya 464-8602, Japan}

\author{H.~Otono}
\affiliation{Department of Physics, Fuculty of Sciences, Kyushu University, Fukuoka 819-0395, Japan}

\author{Y.~Seki}
\affiliation{Institute of Multidisciplinary Research for Advanced Materials, Tohoku University, Sendai 980-8577, Japan}

\author{D.~Sekiba}
\affiliation{Institute of Applied Physics, University of Tsukuba, Tsukuba 305-8573, Japan}

\author{T.~Shima}
\affiliation{Research Center for Nuclear Physics, Osaka University, Ibaraki 567-0047, Japan}

\author{H.~E.~Shimizu}
\affiliation{SOKENDAI, Shonan Village, Hayama 240-0193, Japan}

\author{H.~M.~Shimizu}
\affiliation{Department of Physics, Graduate School of Science, Nagoya University, Nagoya 464-8602, Japan}

\author{N.~Sumi}
\affiliation{High Energy Accelerator Research Organization, Tsukuba 305-0801, Japan}

\author{H.~Sumino}
\affiliation{Research Center for Advanced Science and Technology, The University of Tokyo, Tokyo 153-0041, Japan}

\author{M.~Tanida}
\affiliation{Department of Physics, Graduate School of Science, Kyushu University, Fukuoka 819-0395, Japan}

\author{H.~Uehara}
\affiliation{Department of Physics, Graduate School of Science, Kyushu University, Fukuoka 819-0395, Japan}

\author{T.~Yamada}
\affiliation{Department of Physics, Graduate School of Science, The University of Tokyo, Tokyo 113-0033, Japan}

\author{S.~Yamashita}
\affiliation{Research and Regional Cooperation Office, Iwate Prefectural University, Takizawa 020-0693, Japan}

\author{K.~Yano}
\affiliation{Department of Physics, Graduate School of Science, Kyushu University, Fukuoka 819-0395, Japan}

\author{T.~Yoshioka}
\affiliation{Research Center for Advanced Particle Physics, Kyushu University, Fukuoka 819-0395, Japan}

\begin{abstract}
The ``neutron lifetime puzzle'' arises from the discrepancy between neutron lifetime measurements obtained using the beam method, which measures decay products, and the bottle method, which measures the disappearance of neutrons.
To resolve this puzzle, we conducted an experiment using a pulsed cold neutron beam at J-PARC. In this experiment, the neutron lifetime is determined from the ratio of neutron decay counts to {\Hecapture} reactions in a gas detector. This experiment belongs to the beam method but differs from previous experiments that measured protons, as it instead detects electrons, enabling measurements with distinct systematic uncertainties. By enlarging the beam transport system and reducing systematic uncertainties, we achieved a fivefold improvement in precision. Analysis of all acquired data yielded a neutron lifetime of $\tau_{\rm n}=877.2~\pm~1.7_{\rm(stat.)}~^{+4.0}_{-3.6}{}_{\rm (sys.)}$~s. This result is consistent with bottle method measurements but exhibits a 2.3$\sigma$ tension with the average value obtained from the proton-detection-based beam method.
\end{abstract}
\maketitle

\textit{Introduction--}
A neutron decays into three particles, a proton, an electron, and an antineutrino via weak interactions. The neutron $\beta$ decay lifetime, {\taun}, is a crucial parameter that determines the neutron-to-proton ratio at the onset of Big Bang nucleosynthesis (BBN)~\cite{mathews2005,chowdhury2024}. The combination of the BBN model and the baryon-to-photon ratio derived from the cosmic microwave background observations~\cite{bennett2013,aghanim2020} provides an accurate prediction of the abundance of light elements, allowing tests of physical phenomena in the early universe. Additionally, the $V_{\rm ud}$ term in the Cabibbo-Kobayashi-Maskawa (CKM) matrix can be determined using $\tau_{\rm n}$ and $\lambda$, which is the ratio of axial-vector to vector coupling constants, $g_A$/$g_V$, independently of nuclear models. 
Revised radiative corrections in 2018~\cite{seng2018} suggested the CKM unitarity violation exceeding 2$\sigma$~\cite{PDG2024}, emphasizing the importance of the measurement of the neutron lifetime.
Precise data on {\taun} is also valuable for testing lattice QCD calculations of $g_A$~\cite{chang2018}.

Neutron lifetime has been measured using two primary methods. The first is the beam method~\cite{byrne1996,yue2013}, where neutron $\beta$ decay products, specifically protons in these references,
are counted relative to the number of incident neutrons, yielding an average lifetime of $\tau_{\rm n}^{\rm beam} = 888.0\,\pm\,2.0$~s. 
The second is the bottle method ~\cite{serebrov2005,pichlmaier2010,steyerl2012,arzumanov2015,serebrov2018,pattie2018,ezhov2018,gonzalez2021}, which measures the disappearance of ultra-cold neutrons~(UCNs) confined in a container over time, producing an average value of $\tau_{\rm n}^{\rm bottle} = 878.4\,\pm\,0.5$~s. The 9.5-s (4.6$\sigma$) discrepancy between the two methods is known as the ``neutron lifetime puzzle''~\cite{greene2016}, raising concerns about the reliability of neutron lifetime measurements.

Possible causes for this discrepancy include unaccounted systematic uncertainties, such as protons from neutron decay undergoing charge exchange with residual gas~\cite{serebrov2021}, though this effect is considered negligible~\cite{wietfeldt2023}.
The 9.5-s, approximately 1\% discrepancy between beam and bottle methods has caused discussions of theories such as UCN inelastic scattering with dark matter~\cite{rajendran2021}, hydrogen atom formation~\cite{oks2024}, mirror neutron transitions~\cite{berezhiani2019}, or neutron decay into dark matter~\cite{fornal2018}. Upcoming experiments~\cite{materne2009,wei2020,auler2024,krivovs2024}, including space neutron detections~\cite{wilson2021,tsuji2023}, may help resolve this issue.

To address the neutron lifetime puzzle, we conducted neutron lifetime measurements using the high-intensity pulsed neutron beamline at Japan Proton Accelerator Research Complex (J-PARC), adopting the method of Kossakowski et al.~\cite{kossakowski1989}.
This experiment detects electrons from neutron $\beta$ decay, offering a distinct observable compared to previous proton-counting beam methods~\cite{byrne1996,yue2013}, and thus provides an alternative approach to verify the neutron lifetime puzzle. Our first result in 2020, $\tau_{\rm n} = 898\,\pm\,10_{\rm (stat.)}\,^{+15}_{-18}\,_{\rm (sys.)}$~s~\cite{hirota2020}, was consistent with both beam and bottle methods. 
Through improved neutron intensity by a new neutron transport and reduced systematic uncertainties, we achieved a fivefold improvement in precision, as discussed in this paper.

\begin{figure}[th]
\centering
\includegraphics[keepaspectratio,width=\linewidth]{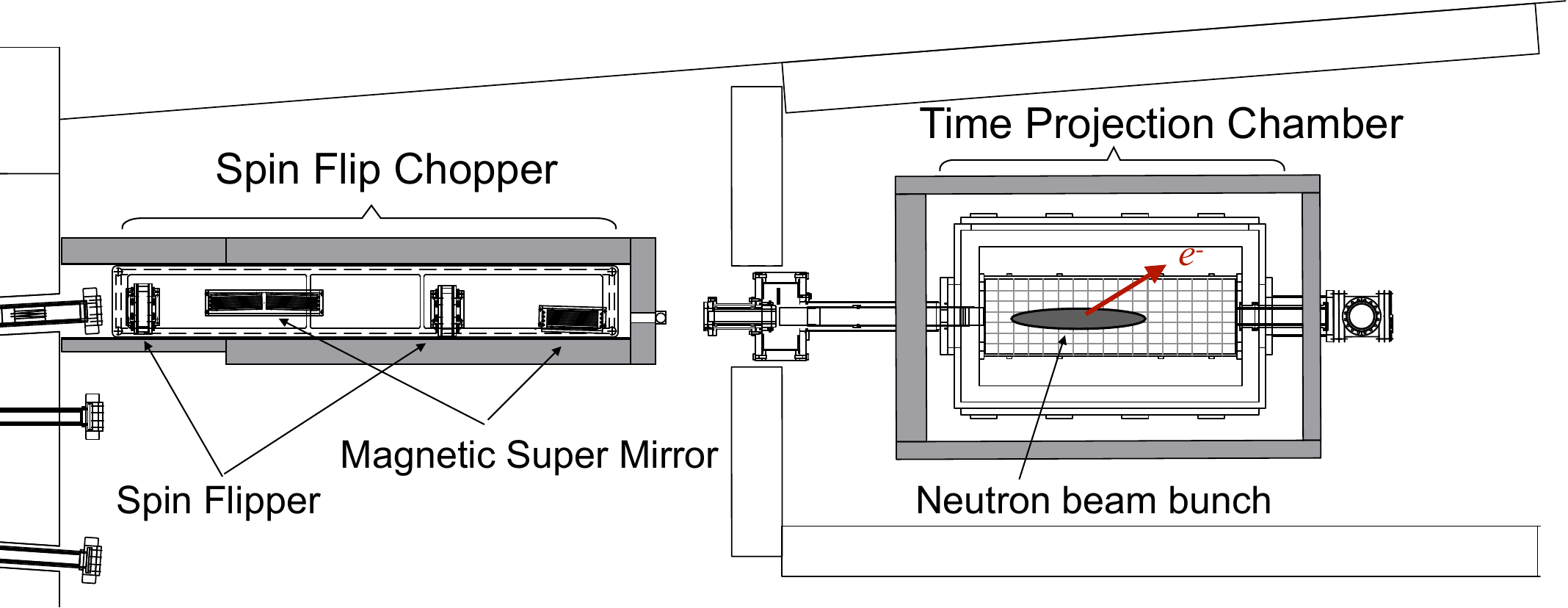}
\caption{
The top view of BL05/NOP. Pulsed neutrons entering from the beam port on the left are shaped into bunches by the spin flip chopper and then directed into the time projection chamber.
}
\label{fig:setup}
\end{figure}

\textit{Principle of the experiment--}
The experiment, conducted at the polarized neutron beam branch of BL05/NOP at J-PARC Materials and Life Science Experimental Facility~\cite{mishima2009,nakajima2017}, is illustrated schematically in Fig.~\ref{fig:setup}.
The pulsed neutron beam is shaped into 40-cm-long bunches using a spin flip chopper (SFC)~\cite{taketani2011,ichikawa2022,mishima2024} and introduced into a 1-m-long time projection chamber (TPC)~\cite{arimoto2015}. Measuring $\beta$ decay events while the neutron bunch is fully within the TPC reduces uncertainties in detection efficiency and effects from the background.

The TPC is filled with an He+CO$_2$ gas with a mixture of 85:15, optimized for low neutron scattering and efficient electron multiplication. This setup achieves a trigger efficiency of over 99.9\% for neutron $\beta$ decay events.
A precise amount of {\HeT}, corresponds to the pressure of 50~mPa, is added using a volumetric expansion method, and {\Hecapture} reactions are used to determine the number of incident neutrons.
The inner walls of the beam duct and TPC are covered with a shielding material made of a mixture of $^{6}$Li-enriched LiF and polytetrafluoroethylene (PTFE) in a 30:70 weight ratio~\cite{koga2021}. This cover effectively reduces background $\gamma$ radiation produced by neutron capture.

The neutron lifetime is derived using the detected $\beta$ decay count $S_{\beta}$, the {\Hecapture} reaction count $S_{\rm He}$, their respective detection efficiencies $\varepsilon_{\beta}$ and $\varepsilon_{\rm He}$, the number density of {\HeT} in the detector $\rho$, and the {\Hecapture} reaction cross-section $\sigma_0$ at the neutron velocity $v_0 = \qty{2200}{m/s}$, according to the following equation~\cite{hirota2020}:

\begin{equation}
\tau_{\rm n} = \frac{1}{\rho\sigma_0 v_0}\left( \frac{S_{\rm He}/\varepsilon_{\rm He}}{S_\beta/\varepsilon_\beta} \right).
\label{eq_tau_0}
\end{equation}

The $\beta$ decay events and {\Hecapture} events are identified using waveform and track information from the TPC. The neutron lifetime is calculated from their count ratio and detection efficiencies, determined by using Monte Carlo simulations (MCs) based on  Geant4.9.6~\cite{allison2016}.

\textit{Updates--}
We conducted experiments with improvements to the apparatus and analysis based on the 2020 results~\cite{hirota2020}. A major source of systematic uncertainties was background $\gamma$ rays from the absorption reactions of neutrons scattered by the TPC working gas. To address this, we reduced the gas pressure to mitigate neutron scattering and its associated effects. Lower pressures made the TPC more prone to electrical discharge, necessitating reduced signal amplification. 
By minimizing electrical noise and optimizing the operating conditions, we successfully obtained stable measurements at 50~kPa, reduced from the original pressure of 100~kPa. In addition, simulations were developed to reproduce these low-pressure conditions, incorporating specific wire efficiency and space charge. These simulations thereby enable reliable measurements at 50~kPa.
The low-pressure measurements further improved the accuracy of determining the {\HeT} content in the working gas and also reduced background events caused by nuclear reactions with CO$_2$, namely {\Ccapture} and {\Ocapture}.

The TPC operating gas used is commercially available high-grade He gas (G1He) containing 0.1~ppm of {\HeT}~\cite{hirota2020}.
Previously, the amount of {\HeT} was determined by analyzing G1He using a mass spectrometer~\cite{sumino2001,mishima2018}. After measuring the {\Ncapture} reaction cross section with a precision of 0.4\%~\cite{kitahara2019}, we established a method to determine the $^3$He/$^4$He ratio in G1He with an accuracy of better than 0.8\% by performing measurements with He+CO$_2$ gas mixed with N$_2$ gas~\cite{mogi2022}.
We also improved the method for determining the amount of {\HeT} introduced into the TPC vacuum vessel. In this experiment, 50~mPa of {\HeT} gas is precisely injected using a buffer vessel with a known volume ratio to the vacuum chamber. By employing two pressure gauges with different dynamic ranges and an optimized buffer volume, the measurement precision of the volume ratio improved from 0.34\% to 0.12\%. Consequently, the accuracy of $\rho$ determination at 50~kPa was enhanced to 0.13\%~\cite{mogi2022}. These methods have been adopted since 2019.

We use an SFC to bunch neutrons incident to the TPC. The neutron intensity was previously limited by the size of the neutron mirrors of the SFC~\cite{taketani2011}. The upgraded SFC, featuring larger magnetic supermirrors and spin flippers, increased the neutron intensity to the TPC by a factor of 2.8 while maintaining a signal-to-noise ratio of 250--400~\cite{taketani2011,ichikawa2022,mishima2024}.
Measurements were performed under four conditions: two gas pressures (100~kPa and 50~kPa) and two SFC configurations (old and new SFC). Table~\ref{table:RunSummary} summarizes the dataset, comprising 49 gas fills measured between 2014 and 2023. Each gas fill represents a measurement of approximately one week, including energy calibration with a $^{55}$Fe source, drift time calibration using cosmic rays, neutron lifetime measurements with beam passage, and background measurements with a beam shutter closed. During the measurement, the beam alternated periodically between a beam pass run (1000--1200~s) and a beam dump run (800--1000~s).

\begin{table}[htbp]
\caption{
Measurements for each condition, presented as the number of gas fills and beam cycles (defined as one beam pass run and one dump run).
}
\label{table:RunSummary}
\centering
\resizebox{0.9\columnwidth}{!}{
\begin{tabular}{cccc}
\hline
Conditions & Year & Gas fill & Beam cycle \\ \hline
100~kPa/old SFC & 2014--2019 & 29 & 4660 \\ 
50~kPa/old SFC & 2017--2018 & 6 & 1044 \\   
100~kPa/new SFC & 2021--2023 & 5 & 671 \\ 
50~kPa/new SFC & 2021--2023 & 9 & 1729 \\ 
\hline
\end{tabular}
}
\end{table}

\begin{figure}[tbhp]
\centering
\includegraphics[keepaspectratio,width=0.6\linewidth]{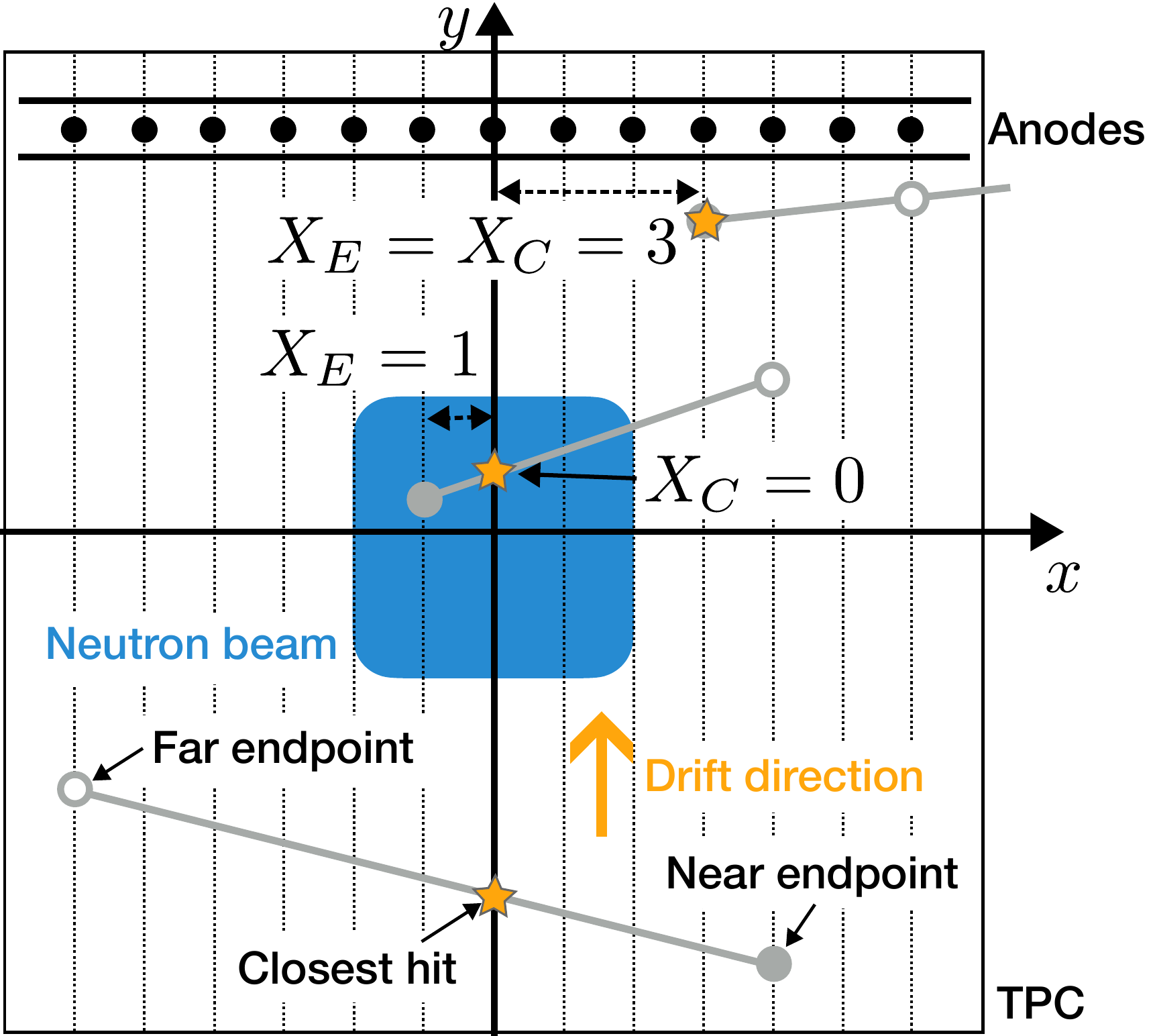}
\caption{Conceptual downstream view of the TPC illustrating $X_E$ (distance from the near endpoint of a track to central wire) and $X_C$ (distance from closest hit to center). Number of wires and beam size are not to scale.
}
\label{fig:Xdef}
\end{figure}

\textit{Analysis--}
The analysis follows the method of previous measurements~\cite{hirota2020}. Two parameters, $X_E$ and $X_C$ are introduced to separate $\beta$ decay events from the background.
Here, $X_E$ is the distance between the near endpoint of a track and the central wire, while $X_C$ indicates the distance from the closest hit to the center, expressed in wire spacing units (Fig.\ref{fig:Xdef}). 
Since neutron $\beta$ decay events originate from the beam region (40~mm), the signal region was defined as $X_E < 5$ ($<$~60~mm). 
On the other hand, the region $X_C \geq 5$ contains almost no events originating from neutron $\beta$ decay and was therefore defined as the background region.

The primary background in this experiment arises from $(\mathrm{n},\gamma)$ events caused by neutrons scattered by the TPC working gas.
The number of background events to be subtracted can be obtained by multiplying the number of events in the background region by the ratio of those entering the signal region to those in the background region. This ratio is determined using MCs.
A larger background than predicted by MC simulations was observed, amounting to 4.9--5.4\% of $S_\beta$ at 100~kPa (vs.~1.2--1.3\% predicted) and 3.1--3.3\% at 50~kPa (vs.~0.65--0.67\%). 
While the cause remains unclear, leakage of scattered neutrons through gaps in the {\LiF} is suspected.
To model this background, we selected MCs assuming discrete $\gamma$ ray energies that best reproduced the experimental results.
Since the single-energy distribution used in the previous method~\cite{hirota2020} could not reproduce the background shape or  the energy loss per unit length ($\mathrm{d}E/\mathrm{d}x$)
distribution, we constructed MCs with $\gamma$ rays with two energy components. 
The $\gamma$ ray energies were determined to reproduce the two-dimensional distribution of $X_E$ and $X_C$ outside the signal region.
The origins of the $\gamma$ rays were considered: ``Internal,'' for neutron absorption within the {\LiF} shielding, and ``Internal-external,'' for absorption within and outside the shielding. The ``Internal-external'' model, with 200~keV contributing \SI{91.9(8)}{\percent} and 5000~keV contributing \SI{8.1(8)}{\percent}, showed the best fit to the experimental data with the $\chi^2$ per degrees of freedom (DOF) of $208.8/202$.

\begin{figure}[tbhp]
\centering
\includegraphics[keepaspectratio,width=\linewidth]{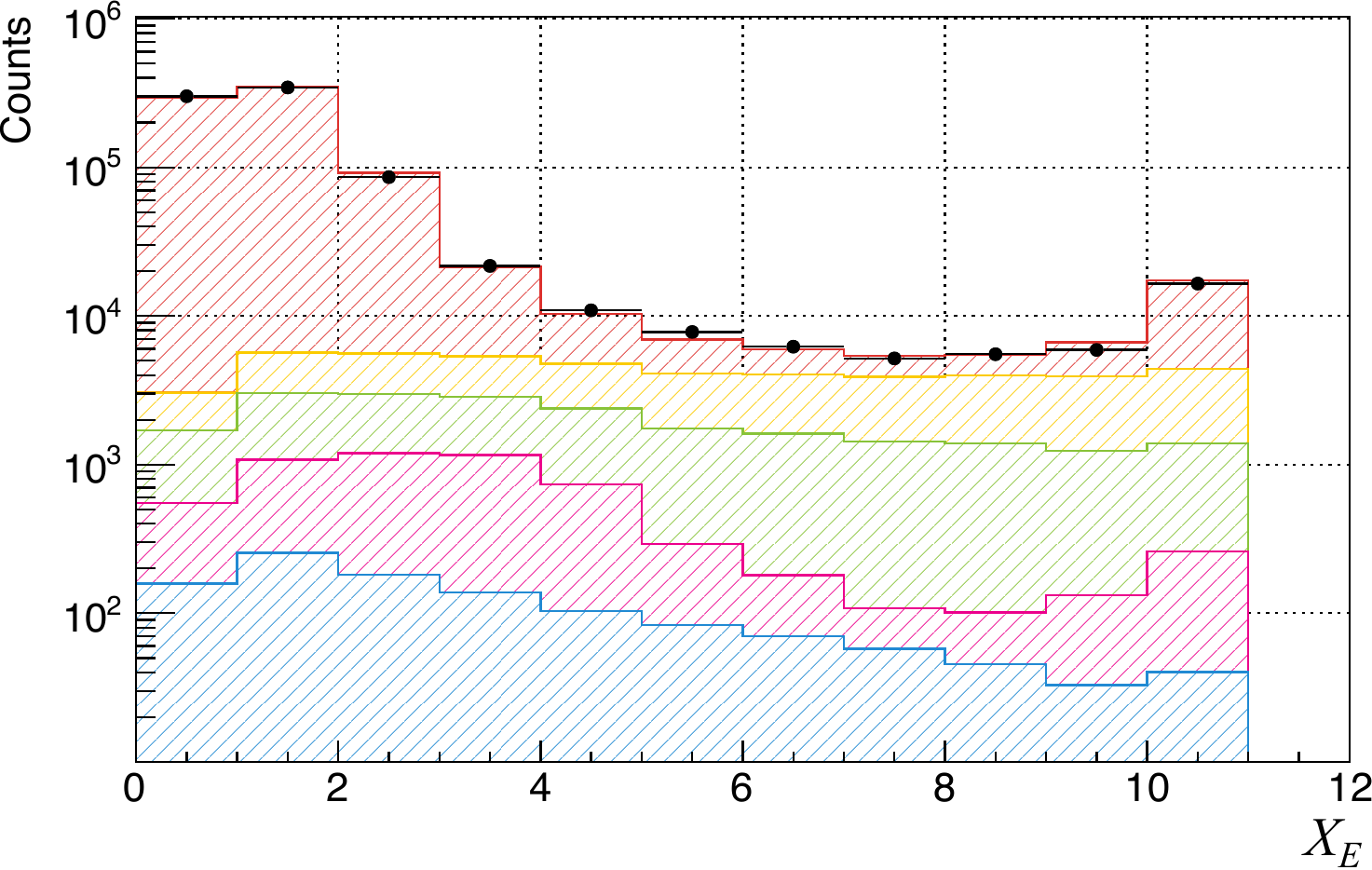}
\caption{
$X_E$ distribution for \qty{50}{kPa}/new SFC. Black points represent measured data, and colored regions represent the stacked MC distributions. From top to bottom: red ($\beta$), yellow (gas-scattered \qty{5000}{keV} $\gamma$ rays), green (gas-scattered \qty{200}{keV} $\gamma$ rays), magenta ($\gamma$ rays scattered by the {\LiF} shutter), and blue ($\beta$ events from gas-scattered neutrons).
}
\label{fig:DC}
\end{figure}

Signal cut uncertainties were evaluated by varying the $X_E$ cut from 4~ch to 11~ch (5~ch as baseline), with a maximum variation of 0.2\% in the data-to-MC event ratio (Fig.~\ref{fig:DC}). Energy cut uncertainties, estimated from $\mathrm{d}E/\mathrm{d}x$ differences for cosmic rays, contributed less than 0.1\%. These improvements ensured accurate background subtraction and enhanced the reliability of neutron $\beta$ decay measurements.

The W-value, the average energy required for single-electron ionization, in the He+CO$_2$ gas of a proton ($W_{\mathrm{p}}$) increases as proton energy decreases below 1~keV unlike electron ionization ($W_{\mathrm{e}}$)~\cite{katsioulas2022}. We previously addressed the effect by simplifying the quenching factor as $q(E) = W_{\mathrm{e}}(E)/W_{\mathrm{p}}(E) = 1$ with kinetic energy $E$, while accounting for the impact of $q(E) = 0$ as a 0.6\% systematic uncertainty. Improved statistics now make this effect significant, thus $q(E)$ value from SRIM~\cite{ziegler2010}, ranging from 0.2 to 0.6 for 25--750~eV protons, were adopted. This adjustment reduced MC detection efficiency $\varepsilon_{\beta}$ by up to 0.3\%, with a systematic uncertainty of 0.1\% based on differences between W-values from SRIM and at which MCs give best fits.

Pileup events were estimated based on count rates. The full pileup rate corresponds to a lifetime change of 0.4\%. We evaluated the classification of primary pileup events, triggered by a $\beta$ decay electron and a {\HeT} event follows, by developing a dedicated MC simulation that generates events at different timings. This improved the classification accuracy, reducing the pileup uncertainty to +0.17/-0.07\%.

\textit{Results--}
To mitigate human bias, a random variation of $\pm$10\% has been applied to $\rho$ during the updated analysis, which was removed only after the analysis method and parameters had been finalized.
Data were grouped into four conditions based on gas pressure (100~kPa and 50~kPa) and SFC type (old and new), with average neutron lifetimes summarized in Table~\ref{table:FourResults}. For each condition, uncertainties related to cut positions were evaluated independently, while globally correlated uncertainties were applied as common shifts. 
The overall neutron lifetime result from J-PARC, combining all conditions, is 
$\tau_{\rm n} = 877.2 \pm 1.7_{\rm(stat.)}~^{+4.0}_{-3.6}{}_{\rm(sys.)}$.
The combining average yielded $\chi^{2}/\mathrm{DOF} = 15.8/3$, though the underlying cause of this deviation remains undetermined.

\begin{table}[tbhp]
\caption{
Neutron lifetime values for each gas pressure (100~kPa, 50~kPa) and SFC configuration (new, old), with averages. Units in seconds.
}
\label{table:FourResults}
\centering
\resizebox{\columnwidth}{!}{
\begin{tabular}{ccccc}
\hline
Conditions & Value & Stat.& Cut position& Other sys. \\ \hline
100~kPa/old SFC & 870.9  &3.5  &+1.8/-2.8 &+5.5/-4.9 \\
100~kPa/new SFC & 868.3  &4.0  &+1.5/-2.9 &+3.8/-3.2\\ 
50~kPa/old SFC  & 868.2  &7.7  &+2.7/-0.9 &+4.8/-3.9 \\
50~kPa/new SFC  & 884.8  &2.4  &+0.8/-1.3 &+3.2/-3.0 \\ \hline
Combined         & 877.2 &1.7   &          & +4.0/-3.6  \\ \hline
\end{tabular}
}
\end{table}

\begin{table}[tbhp]
\caption{
List of uncertainties with units in seconds.
}
\label{table:error_budgets}
\begin{ruledtabular}
\begin{tabular}{cr}
Effect & Uncertainty  \\ \colrule
Statistic& 1.7 \\ \hline
Cut position & 0.9 \\ 
Gas-induced background & +1.1/-2.0 \\ 
Pile up& +1.5/-0.6 \\ 
Contamination from {\Ccapture}& +1.7/-0.0\\ 
$\gamma$-ray scattering at LiF shutter&1.3\\ 
Unbunched neutron from SFC& +1.1/-1.0\\ 
Inject {\HeT} & 1.2\\ 
{\HeT} in G1He & +1.5/-1.4\\ 
{\Hecapture} cross section & 1.2\\ 
Total systematic & +4.0/-3.6\\
\end{tabular}
\end{ruledtabular}
\end{table}

The main uncertainties in this measurement are summarized in Table~\ref{table:error_budgets}, with the largest contribution from the gas-scattering neutron background. The effect due to the cut position uncertainties are small (0.9~s) due to good agreement between MC simulations and experimental results.
The {\Ccapture} reaction, where $^{12}$C in the CO$_2$ gas absorbs neutrons, produces $^{13}$C with a 4946~keV $\gamma$ ray and a 1.0~keV recoil. Misclassification of these events as $\beta$ decay was evaluated using MC simulations, with uncertainties near the energy threshold included in the systematic uncertainty.
The effect due to scattering of $\gamma$ rays at the {\LiF} shutter was modeled using PHITS3.20~\cite{sato2018} and NaI detector measurements. This contributed to a 1.3~s systematic uncertainty. 
The effect of unbunched neutrons caused by SFC imperfections, leading to detection efficiency mismatches for $\beta$ and {\Hecapture} events at the TPC edges, was simulated to evaluate their impact and incorporated as systematic uncertainties.
The uncertainties in introduced {\HeT} and G1He-contained {\HeT} were incorporated into $\rho$. For 50~kPa operations, $^{14}$N-based measurements limited the uncertainty to 0.5~s, while mass spectrometer data increased it to 1.5~s. The {\Hecapture} cross-section, $5333 \pm 7$~barn~\cite{mughabghab2006} obtained by averaging two experiments~\cite{als-nielsen1964, alfimenkov1977}, contributed 1.2~s of uncertainty.

\begin{figure}[tbhp]
\centering
\includegraphics[keepaspectratio,width=\linewidth]
{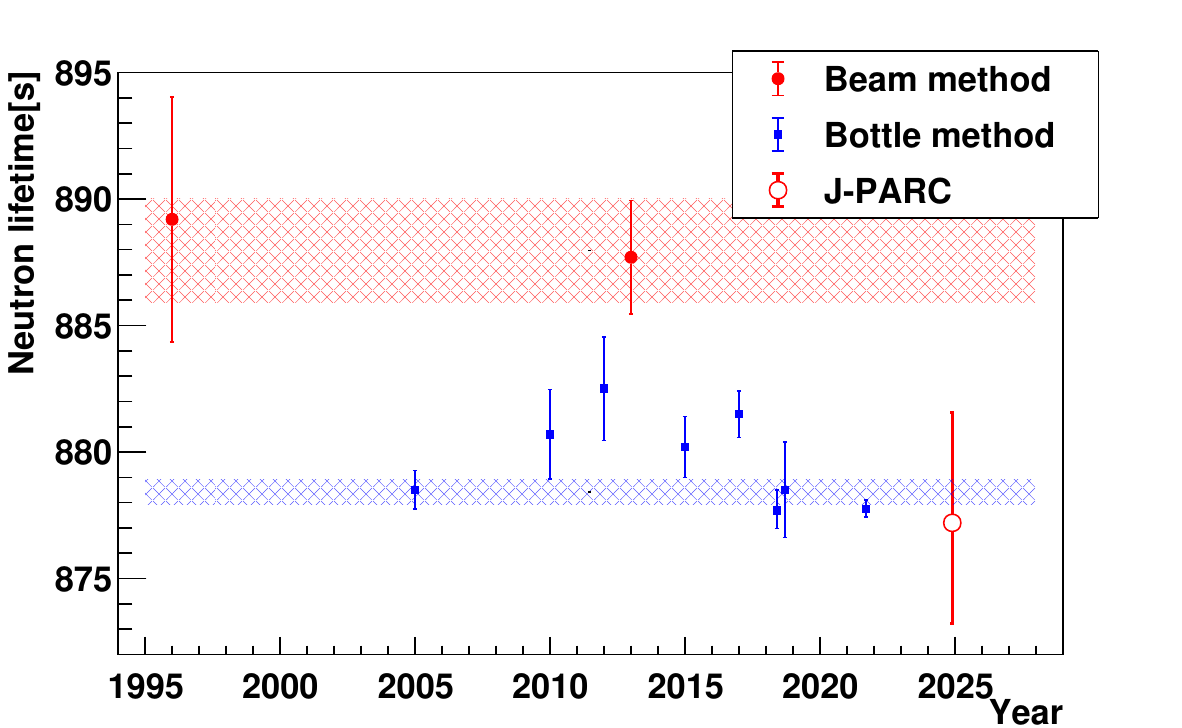}
\caption{
Measured neutron lifetimes in this work and previous experiments~\cite{PDG2024}. Red circles represent measurements using the proton-detection-based beam method, blue squares represent the bottle method, and the red and blue bands indicate the averages of these two methods. The open circle represents the measurement from this work at J-PARC.}
\label{fig:lifetime}
\end{figure}

The neutron lifetime obtained in this study, with statistical and systematic uncertainties combined in quadrature is shown in Fig.\ref{fig:lifetime}, along with results from previous experiments~\cite{PDG2024}.
Our value is consistent with the bottle method but shows a 2.3$\sigma$ tension with the average of the proton-counting beam method. Combining our results with other beam measurements gives $\tau_{\mathrm{n}}^{\rm beam} = 886.0 \pm 1.8$, reducing the discrepancy with the bottle method to 4.0$\sigma$.

\textit{Conclusion and outlook--}
To address the ``neutron lifetime puzzle,'' arising from discrepancies between neutron lifetimes measured by the beam method (via decay products) and the bottle method (via disappearance), we measured the neutron lifetime from the ratio of electrons from neutron decay to {\Hecapture} reactions. Unlike previous beam method experiments that detected protons, this experiment introduced distinct systematic uncertainties.

By enlarging the SFC aperture, the neutron intensity increased by a factor of 2.8, enabling high-statistics data acquisition and achieving a statistical precision of 1.7~s. Systematic uncertainties were reduced to 4~s by accurately modeling gas-scattered neutron backgrounds. Combining all data, the neutron lifetime was determined as $\tau_{\rm n}=877.2~\pm~1.7_{\rm(stat.)}~^{+4.0}_{-3.6}{}_{\rm (sys.)}$~s. This result aligns with the bottle method and is 10.8~s shorter than the averaged value of proton-counting beam method, exhibiting 2.3$\sigma$ tension in the results of the beam method.

The primary limitation of this experiment is the $\gamma$ ray background from gas-scattered neutrons in the TPC. To suppress this background, the upcoming LiNA experiment~\cite{otono2017,sumi2023} will apply a solenoidal magnetic field to the TPC. This setup is expected to reduce background events by a factor of 50. Furthermore, the reduction of background will decrease the required measurement time to one-third, while the uncertainty due to pileup is also anticipated to decrease to one-third as a result of its shorter drift length.

\begin{acknowledgments}
We sincerely thank Prof.~Y.~Namito and H.~Hirayama for the discussions on Geant4 specifications, and Prof.~M.~Hino for fabricating the neutron supermirror for the SFC.
This work was supported by JSPS KAKENHI Grant Numbers (JP19GS0210, 23244047, 24654058, 26247035, 16H02194, 18H01231, 19H00690, 21H04475, and 22H00140), and
JST, the establishment of university fellowships towards the creation of science technology innovation, Grant Number JPMJFS2136.
The neutron experiment at the Materials and Life Science Experimental Facility of J-PARC was performed under user programs (Proposal Nos. 
2014B0271, 2015A0316, 2016A0168, 2016B0272, 2017A0230, 2017B0332, 2018A0297, 2018B0272, 2019A0223, 2019B0341, 2020A0223, 2021B0287, 2022A0117, 2022B0325, 2023A0184, 2023B0253, and 2024A0331) and an S-type projects of KEK IMSS (Proposal Nos. 2014S03 and 2019S03). This experiment is supported as stage-1 status (E100) of J-PARC-PAC KEK IPNS.
\end{acknowledgments}

\bibliography{Ref}
\end{document}